\documentclass[10pt,letterpaper]{article}

\usepackage[margin=1in]{geometry}
\usepackage[T1]{fontenc}
\usepackage{lmodern}
\usepackage[utf8]{inputenc}
\usepackage[protrusion=true,expansion=false]{microtype}
\usepackage{graphicx}
\usepackage{booktabs}
\usepackage{amsmath,amssymb}
\usepackage[numbers,sort&compress]{natbib}
\usepackage[strings]{underscore}
\usepackage{xspace}
\usepackage[section]{placeins}
\usepackage[hidelinks]{hyperref}

\title{Spark Policy Toolkit: Semantic Contracts and Scalable Execution for Policy Learning in Spark}
\author{Zeyu Bai\\\texttt{zeyubai21@engineering.ucla.edu}}
\date{}

\newcommand{\N}{\ensuremath{N}\xspace}
\newcommand{\F}{\ensuremath{F}\xspace}

\newcommand{\T}{\ensuremath{T}\xspace}

\newcommand{\pmiss}{\ensuremath{p_{\mathrm{miss}}}\xspace}

\begin{document}
\maketitle

\begin{abstract}
Custom policy-learning pipelines in Spark fail for two coupled systems reasons:
rowwise Python execution makes inference impractical, and driver-side candidate
materialization makes split search fragile at feature scale. We present Spark
Policy Toolkit, a semantics-governed systems toolkit for scalable policy
learning in Spark. The toolkit provides two Spark-native primitives:
partition-initialized vectorized inference through \texttt{mapInPandas} and
\texttt{mapInArrow}, and collect-less split search that scores candidates on
executors. Both primitives are governed by one fixed-input semantic contract:
the same rows, feature order, treatment vocabulary, preprocessing manifest, and
split boundaries must preserve per-row score vectors, best-split decisions, and
end-to-end learned policy outputs. The evaluation combines practical baseline
ladders, backend parity checks, measured split-search scale results, synthetic
and Hillstrom end-to-end policy preservation, missingness stress, partition and
order perturbation tests, quantile-boundary sensitivity, and a concrete
adversarial failure catalog. On a 40-worker Databricks cluster,
\texttt{mapInArrow} reaches 4.72M rows/s at 10M matched rows and 7.23M rows/s
at 50M rows, while collect-less split search remains valid from
\(\F=10\) through \(\F=1000\) with 124000 candidate rows, where the
driver-collect baseline is intentionally skipped. Across 24 backend-ablation
settings, \texttt{mapInArrow} wins 18 while \texttt{mapInPandas} wins 6, so the
paper treats backend choice as workload-dependent rather than universal. Once
the fixed-input lock is enforced, all six tested repartition/coalesce/shuffle
perturbations preserve identical signatures; before lock, all six drift. The
central result is not speed alone: throughput and collect-less execution are
the mechanisms that let policy semantics survive at Spark scale.
\end{abstract}

\section{Introduction}
Policy and uplift workflows often sit between several adjacent literatures.
Distributed tree systems such as PLANET, XGBoost, LightGBM, and Spark MLlib
optimize standard supervised objectives
\cite{panda2009planet,chen2016xgboost,ke2017lightgbm,meng2016mllib}, while
uplift and heterogeneous-effect libraries often target local or model-centric
workflows
\cite{athey2016causal_tree,wager2018causal_forest,rzepakowski2010uplift_trees,chen2020causalml,scikituplift_github}.
Distributed causal infrastructure has also become more practical, including
distributed double machine learning, Spark-oriented causal APIs such as
SynapseML's \texttt{DoubleMLEstimator}, and production uplift systems such as
H2O's distributed uplift random forest
\cite{kurz2021doubleml_serverless,hamilton2018mmlspark,synapseml_causal_overview,h2o_upliftdrf}.
At the same time, a separate correctness lineage asks how to make machine
learning execution deterministic or semantically faithful, from deterministic
workflow frameworks such as \texttt{mlf-core} to graph-equivalence verification
systems such as AERIFY \cite{heumos2023mlfcore,zulkifli2025aerify}. Production
Spark teams still build custom policy pipelines around application-specific
treatment vocabularies, missing-value rules, split objectives, and business
utility witnesses. These pipelines are easy to express in Python and easy to
break when distributed execution is introduced.

The systems failure is coupled. Inference commonly falls back to per-row Python
model deserialization or rowwise scoring. Split search commonly collects
candidate statistics to the driver, where candidate count grows with features,
bins, treatments, and missing-routing variants. Replacing these paths with
distributed operators is not a mechanical optimization: small changes in
missing routing, control assignment, tie resolution, candidate validity, or
bucket boundaries can silently change the learned policy.

This work therefore targets a gap not fully addressed by standard distributed
tree systems, distributed causal estimators, or local uplift packages. It does
not introduce a new causal or uplift objective. It is also narrower than a
general model-equivalence verifier such as AERIFY: the target here is not
arbitrary tensor-program equivalence, but exact preservation of row-level policy
vectors, split semantics, and shallow policy-tree witnesses for an existing
Spark policy pipeline. Instead, the paper addresses the systems pathologies that
appear when custom policy-learning code is embedded inside Spark jobs with
Python-side execution and ad hoc distributed rewrites
\cite{zaharia2012rdd,armbrust2015sparksql,saur2022containerized_udfs}. The
central question is how to make an existing custom policy pipeline scalable
without changing what that pipeline means.

This paper therefore frames Spark Policy Toolkit as a semantics-governed
execution layer, not as a pure speed paper and not as a general determinism
paper. Throughput and collect-less execution are not a separate speed story;
they are the enabling mechanisms that let the semantic contract survive at
production scale.

The contributions are:
\begin{itemize}
\item a Spark-native vectorized inference primitive with \texttt{mapInPandas}
and \texttt{mapInArrow} execution paths;
\item a collect-less split-search primitive that scores candidates without
driver-side candidate-table materialization;
\item a fixed-input semantic contract covering data, inference, split search,
and end-to-end trainer witnesses;
\item an experiment program that tests performance, parity, policy
preservation, missingness stress, partition/order robustness, boundary
sensitivity, and concrete adversarial failures.
\end{itemize}

\section{Problem and Threat Model}
The target workload is a custom multi-treatment policy-learning pipeline in
Spark. A trained policy forest or tree assigns each row a treatment-score
vector, and split search chooses candidate thresholds from per-treatment outcome
statistics. The threat is not malicious input. The threat is a distributed
rewrite that appears equivalent but changes one of the semantics that determines
the final policy.

We consider five concrete drift mechanisms. First, implicit null handling can
route missing feature values differently from the learned node rule. Second,
first-seen control selection can make treatment identity depend on partition or
row order. Third, unstable argmax or sort behavior can change ties between equal
candidate scores. Fourth, sparse-group omission can treat absent
bin-by-treatment combinations as if they did not constrain validity. Fifth,
independent approximate quantile recomputation can change bucket boundaries,
which changes candidate membership before the split score is evaluated.

The claims in this paper are fixed-input claims. They do not assert general
determinism under arbitrary Spark conditions, arbitrary preprocessing, or
independently recomputed approximate boundaries.

\section{Toolkit Overview}
Spark Policy Toolkit exposes two primitives that share one semantic surface.
The first primitive scores array-native policy trees over Spark partitions. A
forest is broadcast once, vectorized tree objects are initialized once per
partition, and batches are scored through \texttt{mapInPandas} or
\texttt{mapInArrow}. A broadcast-rowwise path remains as a practical comparator,
and an intentionally slow JSON-per-row path remains as a legacy anti-pattern
baseline.

The second primitive performs collect-less split search. Spark bucketization
assigns feature values, including an explicit missing bin. Executor-side
aggregation builds per-bin, per-treatment prefix sums. Candidate scoring then
evaluates both NaN-left and NaN-right routes and applies a deterministic
score-and-tie order. The driver receives only the selected best-split tuple,
not the full candidate table.

\subsection{Primitive-Level Data Flow}
At the API level, the toolkit accepts ordinary Spark DataFrames together with a
small set of explicit policy-learning inputs. Primitive A takes feature columns,
an ordered feature list, and a forest represented as array-native trees, then
returns one Spark array column containing the per-row treatment-score vector.
Primitive B takes a feature column, treatment labels, outcomes, and fixed split
boundaries, then returns a deterministic best-split object with feature,
candidate bin, threshold boundary, missing-value direction, score, and optional
diagnostics. A driver-collect implementation remains available only as a
reference path for parity and small-scale comparisons.

\paragraph{Primitive A.}
Inputs are a Spark DataFrame, an ordered \texttt{feature\_cols} list aligned to
the tree schema, and a forest represented as array-native trees. The output is a
new DataFrame with one appended Spark array column containing the length-\T
treatment-score vector for each row. The production backends are
\texttt{mapInPandas} and \texttt{mapInArrow}; \texttt{broadcast\_rowwise} and
\texttt{anti\_pattern} exist only as comparators.

\paragraph{Primitive B.}
Split search follows a two-stage contract. Stage S1,
\texttt{build\_prefix\_sums(...)}, converts a DataFrame containing
\texttt{feature\_col}, \texttt{treatment\_col}, and \texttt{outcome\_col} into a
fixed prefix-sum table using explicit boundaries. Stage S2,
\texttt{score\_candidates\_collectless(...)}, takes that prefix table plus
constraints such as \texttt{min\_leaf\_size} and returns a deterministic
\texttt{BestSplit} record. The production execution path is collect-less; the
driver-collect version is retained only as a reference implementation.

\subsection{Execution Contract}
The two primitives are coupled by design. Inference is not only a deployment
path; it is also the mechanism that turns a serialized policy model into a
stable row-level policy-vector witness. Split search is not only a training
optimization; it is the mechanism that ensures the same bucketized statistics
lead to the same candidate-validity set and the same winning split under
distributed execution. That is why the paper treats them as one toolkit rather
than two unrelated accelerators.

\section{Fixed-Input Semantic Contracts}
The toolkit is governed by four contracts.

\paragraph{C1: Data contract.}
The compared execution paths must use a fixed row set, fixed row identifiers,
fixed feature order, fixed treatment vocabulary, fixed deterministic
preprocessing manifest, and fixed bucket boundaries when split search is
compared.

\paragraph{C2: Inference contract.}
The same forest arrays and the same input rows must produce the same per-row
treatment-score vectors across broadcast-rowwise, \texttt{mapInPandas}, and
\texttt{mapInArrow} paths, up to the declared floating-point tolerance.

\paragraph{C3: Split-search contract.}
The same input table and the same bucket boundaries must produce the same
control identifier, candidate-validity set, best-split tuple, NaN direction, and
score across driver-reference, SQL, and executor-local \texttt{mapInPandas}
scoring.

\paragraph{C4: End-to-end trainer witness contract.}
The same preprocessing manifest, shared boundaries, deterministic node
expansion order, and training configuration must produce the same serialized
tree signature and the same held-out policy outputs.

The scope limits are explicit. Independently recomputed approximate boundaries
can break semantics. Row-order, feature-order, or vocabulary drift outside the
manifest is not covered. Claims are for explicit fixed-input contracts, not for
arbitrary distributed conditions.

Operationally, C2 fixes traversal semantics as part of the contract. Continuous
nodes route left on \texttt{<=}; categorical nodes route left on equality;
missing values follow the per-node \texttt{nan\_goes\_left} flag; and forest
output is the arithmetic mean of the per-tree score vectors. C3 similarly fixes
the semantics of control selection and candidate ordering. Control selection
prefers a treatment label equal to \texttt{control} (case-insensitive), then an
exact label \texttt{"0"}, then the lexicographically smallest label. Candidate
ordering is total: higher score, then lower threshold boundary, then lower
candidate bin, then preferred NaN direction, then feature name.

The C4 witness uses an explicit serialized tree signature rather than only
aggregate metrics. Internal-node records include the node path, split feature,
threshold, candidate bin, and missing-value direction; leaf records include the
node path, deterministic treatment order, and full policy vector. The end-to-end
claim is therefore about the learned tree itself, not only about one utility
number.

\section{Primitive A: Vectorized Inference}
The vectorized inference primitive removes repeated row-level model setup. Each
partition receives the same serialized forest arrays. The \texttt{mapInPandas}
backend converts feature columns into NumPy arrays from Pandas batches; the
\texttt{mapInArrow} backend converts Arrow record batches directly. Both call
the same vectorized tree traversal code and emit one array-valued score column.

The semantic obligation is simple: vectorization must not change score vectors.
The C1 parity block fixes rows, feature order, and forest arrays, then compares
broadcast-rowwise, \texttt{mapInPandas}, and \texttt{mapInArrow} by checksum,
per-row mismatch count, and maximum absolute difference.

\subsection{Why Per-row Execution Fails}
A common legacy deployment pattern in PySpark is to treat the model as row-level
metadata: each row carries a JSON blob describing the forest, scoring parses
that JSON in Python, reconstructs model objects, and then walks the tree one
row at a time. The intentionally slow \texttt{anti\_pattern} baseline captures
that workflow directly. It is slow for structural reasons: JSON decoding and
object construction scale with \N even though the model is constant, rowwise
Python loops defeat NumPy-level vectorization, and transient object churn raises
executor overhead and variability.

To separate worst-case legacy behavior from a more credible practical baseline,
we also evaluate a broadcast-once rowwise scorer
(\texttt{broadcast\_rowwise}). That path broadcasts the forest once,
initializes in-memory model state once per partition, and still scores one row
at a time in Python. The anti-pattern remains the worst-case reference;
practical inference claims in the paper are tied to this stronger rowwise
comparator.

\subsection{Array-native Tree Schema}
The forest is represented as a dictionary of parallel arrays plus a leaf payload
matrix. Each tree stores node type, feature index, threshold or categorical
value, left/right child indices, missing-routing flags, and a leaf response-rate
matrix indexed by node. This representation keeps the broadcast compact and
removes per-row model metadata. It also lets traversal operate on arrays of
current node indices, so each iteration updates the still-active rows in a
vectorized way until all rows reach leaves.

Schema validation is part of correctness rather than hygiene. The implementation
checks required fields, array lengths, child indices, feature indices, and a
traversal step budget that detects malformed trees such as cycles. Those checks
are necessary because a distributed execution path can otherwise fail silently in
ways that resemble benign floating-point noise.

\subsection{Backends and Partition Initialization}
Both supported inference backends are iterator UDFs, but they differ in the
batch representation that Spark hands to Python. \texttt{mapInPandas} receives
Pandas batches, converts the ordered feature columns into NumPy arrays, and
returns one array-valued score column. \texttt{mapInArrow} receives Arrow
\texttt{RecordBatch} objects, extracts features by position, and returns an
Arrow list array. In both cases, the critical design point is lazy
per-partition initialization: the broadcast forest is converted into in-memory
\texttt{VectorizedArrayTree} objects once on the first batch seen by the
partition iterator.

The backend choice is therefore a systems tradeoff, not a semantic one. Arrow
and Pandas can differ in throughput because of interchange and output-materialization
costs, but they are required to execute the same routing semantics under the
fixed-input contract.

\section{Primitive B: Collect-less Split Search}
The split-search primitive separates candidate-stat construction from candidate
scoring. Prefix sums are constructed with explicit treatment levels and explicit
missing-bin counts. Candidate scoring expands each candidate into two missing
routes, validates per-treatment support, computes the multi-treatment DDP
max-envelope score, and selects the winner with a total order:
higher score, lower boundary, lower candidate bin, preferred NaN direction, and
feature name.

The C2 parity block fixes bucket boundaries and compares driver-collect, SQL,
and \texttt{mapInPandas} candidate scoring. Driver-collect is retained only as a
reference and small-scale comparator; the production path is collect-less.

\subsection{Why Driver Collect Fails}
The driver-collect bottleneck has two coupled forms. First, the candidate table
scales as \(O(F \cdot (B-1) \cdot T)\), so even when executors compute the
underlying statistics efficiently, materializing those rows on the driver
creates a single-node memory and Python-loop bottleneck. Second, once the
candidate table is local, the driver becomes the throughput limiter even when it
does not run out of memory. In the validated harness, oversized cases are
safety-skipped before OOM rather than being allowed to fail catastrophically.

\subsection{Bucketization and Prefix Sums}
Stage S1 of the split-search path builds a fixed intermediate representation
with Spark-native operators only. A continuous feature is discretized with
\texttt{Bucketizer(handleInvalid="keep")} \cite{spark_bucketizer}, which turns
NULLs and NaNs into an explicit missing bin. The bucketed data is then grouped
by candidate bin and treatment to compute opportunities and accepts, and missing
bin tallies are carried separately. Window prefix sums
\cite{leis2015efficient_window_functions,spark_sql_window_functions} convert
those tallies into left-branch statistics; right-branch statistics are derived
from totals minus prefixes. The result is a narrow per-candidate sufficient
statistic table that can be reused by multiple scoring backends.

\subsection{Candidate Scoring and DDP Max-envelope}
Stage S2 expands each candidate into both missing-value routes, computes
branch-level response rates, and evaluates the multi-treatment DDP max-envelope
score \cite{rzepakowski2010uplift_trees}. For branch \(b \in \{L,R\}\) and
non-control treatment \(t\),
\[
r_t^{(b)} = \frac{\mathrm{accepts}_t^{(b)}}{\mathrm{opps}_t^{(b)}},
\qquad
u_t^{(b)} = r_t^{(b)} - r_{\mathrm{control}}^{(b)}.
\]
The candidate score is the maximum separation between the best treatment uplift
on one side and the worst treatment uplift on the other:
\[
\mathrm{DDP}_{\max} =
\max \left(
u_{\max}^{(R)} - u_{\min}^{(L)},
u_{\max}^{(L)} - u_{\min}^{(R)}
\right).
\]
The same calculation is implemented in a pure SQL backend and an executor-local
\texttt{mapInPandas} backend. The paper therefore separates two questions:
whether the backends agree semantically, and how their runtimes differ at a
given workload size.

\subsection{Deterministic Validity and Selection}
Candidate validity is part of the contract, not a post-processing detail. A
candidate is valid only if every treatment has positive support on both sides,
total left/right counts satisfy \texttt{min\_leaf\_size}, a control group and at
least one non-control group exist, and the score is defined. The missing-value
direction is optimized explicitly by evaluating both left and right routes for
every candidate. The winner is then selected with the total order described in
Section~4. This is what makes the split-search contract meaningful: the SQL,
\texttt{mapInPandas}, and driver-reference paths are not only trying to return a
similar score; they are required to agree on which candidates are even legal.

\section{Adversarial Failure Modes}
The F1 validation suite includes a concrete failure catalog. The primary
synthetic grid spans row counts 500000 and 2000000, treatment counts 4 and 8,
32 bins, four feature patterns, missing rates 0.0 and 0.3, and both balanced
and severe treatment skew. For 4 treatments, the severe skew vector is 0.90,
0.08, 0.015, 0.005. For 8 treatments, it is 0.88, 0.06, 0.025, 0.015, 0.008,
0.006, 0.004, 0.002.

The feature families are designed to trigger exact or near ties
(\texttt{x_tie}), sparse bin-by-treatment support (\texttt{x_sparse}),
minority-arm missingness (\texttt{x_miss}), ambiguous control labels
(\texttt{x_control}), and near-boundary values (\texttt{x_boundary}). The
contract-preserving path is compared against intentionally stripped-down
variants: no total order, first-seen control, sparse-group omission, implicit
missing routing, and independently recomputed quantiles. A failure mode remains
in the paper only if the experiment records an actual drift.

\begin{table}[tbp]
\centering
\caption{Observed adversarial failures for intentionally naive distributed variants. Some rows fail by selection drift at equal score, so the chosen candidate can change even when the score delta remains zero.}
\label{tab:failure-catalog}
\small
\resizebox{\linewidth}{!}{%
\begin{tabular}{lllll}
\toprule
failure mode & cases & affected features & largest score |delta| & contract rule \\
\midrule
unstable argmax & 29 drift & x\_boundary, x\_miss, x\_tie & 0.000000 (equal-score tie) & deterministic total candidate order \\
first-seen control & 34 drift & x\_boundary, x\_miss, x\_tie & 0.266012 & explicit control priority \\
sparse-group omission & 30 drift, 30 invalid accepted & x\_boundary, x\_miss, x\_sparse, x\_tie & candidate invalid & zero-fill plus full D2.2 validity checks \\
implicit missing routing & 3 drift & x\_miss & 0.116337 & evaluate both missing directions explicitly \\
independent quantile recomputation & 1 invalid candidate set & x\_boundary & candidate invalid & shared fixed boundaries \\
\bottomrule
\end{tabular}
}
\end{table}

\section{Experimental Methodology}
All mandatory experiments are run by
\texttt{databricks\_semantic\_contract\_experiments.py} on Databricks Runtime
15.4 LTS / Spark 3.5.x using 40 fixed workers with autoscaling, Photon, and
spot workers disabled. The mandatory blocks are:
\begin{itemize}
\item P1: practical inference baseline ladder, including matched rows and
production maxima;
\item P2: measured split-search feature scale with no projected main-paper
points;
\item C1: inference backend parity;
\item C2: split-search backend parity;
\item E1: synthetic end-to-end trainer preservation across depths;
\item E2: Hillstrom end-to-end policy parity and utility preservation;
\item F1: adversarial failure catalog;
\item F2: fixed-boundary versus recomputed-boundary sensitivity;
\item F3: missingness stress grid for D1, D2, and a representative end-to-end witness;
\item S1: partition/order perturbation robustness before and after manifest lock;
\item S2: Arrow-vs-Pandas backend crossover and batch-size ablation.
\end{itemize}

The Hillstrom block requires a deterministic preprocessing manifest, fixed row
IDs, fixed feature order, fixed treatment mapping, fixed bucket boundaries, and
a fixed seed/configuration. It reports serialized tree signatures, policy-vector
equality, top-treatment assignment agreement, held-out policy value, AUUC, and
Qini for the same learner family across reference and distributed paths. The
validated manifest contains 64000 source rows, 21 canonical features, and a
deterministic 51200/12800 train/holdout split.

All synthetic blocks use seed 7 and fixed data-generation logic. To keep the
paper synchronized with the final validated evidence, the manuscript cites a
frozen base semantic bundle plus one additive final-round bundle. The frozen
bundle supplies P1/C1/C2/E2 together with the original E1/F2 rows and the
focused P2/F1 rerun; the additive bundle supplies the strengthened E1/F2 rows
plus F3/S1/S2. Tables and figures report only those final validated outputs
rather than intermediate or superseded runs.
Appendix-only legacy-support artifacts are generated separately from the older
\texttt{databricks\_paper\_experiments.py} result family and from the refreshed
8-worker/40-worker E1/E2 reruns. Those appendix artifacts are used only for
secondary systems context and do not replace the semantic-contract evidence.

\begin{table}[tbp]
\centering
\caption{Cluster and Spark/Arrow configuration for regenerated semantic-contract runs.}
\label{tab:config}
\small
\begin{tabular}{p{0.24\linewidth}p{0.66\linewidth}}
\toprule
field & value \\
\midrule
runtime & Databricks Runtime 15.4 LTS / Spark 3.5.x \\
cluster size & 40 fixed workers \\
autoscaling & off \\
Photon & off \\
spot workers & off \\
seed & 7 \\
effective block sources & frozen base: P1/C1/C2/E2; focused rerun: P2/F1; final-round reruns: E1/F2/F3/S1/S2 \\
Hillstrom source rows & 64,000 \\
Hillstrom train / holdout & 51,200 / 12,800 \\
Hillstrom canonical features & 21 \\
Hillstrom outcome & visit \\
\bottomrule
\end{tabular}
\end{table}

\begin{table}[tbp]
\centering
\caption{Validated workload summary for the mandatory semantic-contract blocks.}
\label{tab:workloads}
\small
\begin{tabular}{p{0.07\linewidth}p{0.31\linewidth}p{0.34\linewidth}p{0.18\linewidth}}
\toprule
block & varied factor(s) & fixed core settings & claim tested \\
\midrule
P1 & methods plus matched rows at 1M and 10M; production maxima & F=32, T=4, 50 trees, depth 7 & practical inference ladder \\
P2 & F in \{10, 50, 250, 1000\} & N=100000, T=4, B=32 & measured split-search scale \\
C1 & broadcast\_rowwise, mapInPandas, mapInArrow & fixed rows, feature order, forest arrays & inference parity \\
C2 & driver\_collect, sql, mapInPandas & fixed prefix sums and boundaries & split-search parity \\
E1 & max\_depth in \{1,2,3,4\} & synthetic deterministic policy-tree witness & end-to-end preservation \\
E2 & driver\_collect, sql, mapInPandas & Hillstrom manifest, fixed boundaries, fixed seed & utility-preserving parity \\
F1 & N in \{500k, 2M\}, T in \{4, 8\}, pmiss in \{0.0, 0.3\}, balanced and severe skew & B=32, adversarial features x\_tie/x\_sparse/x\_miss/x\_control/x\_boundary & concrete failure catalog \\
F2 & fixed boundaries versus recomputed approximate quantiles & same data, seed, and backend family & boundary-sensitivity witness \\
F3 & pmiss in \{0.0, 0.1, 0.3, 0.5\}; NULL and NaN; focused missingness & representative D1, D2, and end-to-end witness subsets & missingness stress exactness \\
S1 & repartition/coalesce; shuffle before lock; intra-partition sort on/off & same synthetic population and stable evaluation holdout & partition/order perturbation robustness \\
S2 & batch size, depth, tree count, treatment cardinality & N=5M, F=32, mapInPandas vs mapInArrow & backend crossover / batch-size ablation \\
\bottomrule
\end{tabular}
\end{table}

\section{Results}
\subsection{Performance}
P1 reports two inference views. The matched-row view compares feasible methods
at the same row counts, with the anti-pattern retained only where it is
practical to run. The production-max view reports method-specific upper-scale
runs separately. The practical comparator is broadcast-rowwise; the
anti-pattern remains a worst-case legacy baseline and is not the headline
identity of the paper.

At 1M matched rows, broadcast-rowwise reaches 10522 rows/s, while
\texttt{mapInPandas} and \texttt{mapInArrow} reach 1.12M and 1.07M rows/s,
respectively. At 10M matched rows, broadcast-rowwise remains near 10646 rows/s,
while \texttt{mapInPandas} reaches 3.10M rows/s and \texttt{mapInArrow}
reaches 4.72M rows/s, corresponding to roughly \(291\times\) and
\(443\times\) speedups over the practical rowwise baseline. The production-max
view reaches 6.86M rows/s for \texttt{mapInPandas} and 7.23M rows/s for
\texttt{mapInArrow} at 50M rows. Arrow is therefore not a universal winner: at
1M matched rows, \texttt{mapInPandas} is slightly faster, while Arrow overtakes
it by 10M rows and stays ahead at the 50M production maximum. That is exactly
the kind of backend-choice story the paper should tell.

\begin{figure}[tbp]
\centering
\includegraphics[width=0.95\linewidth]{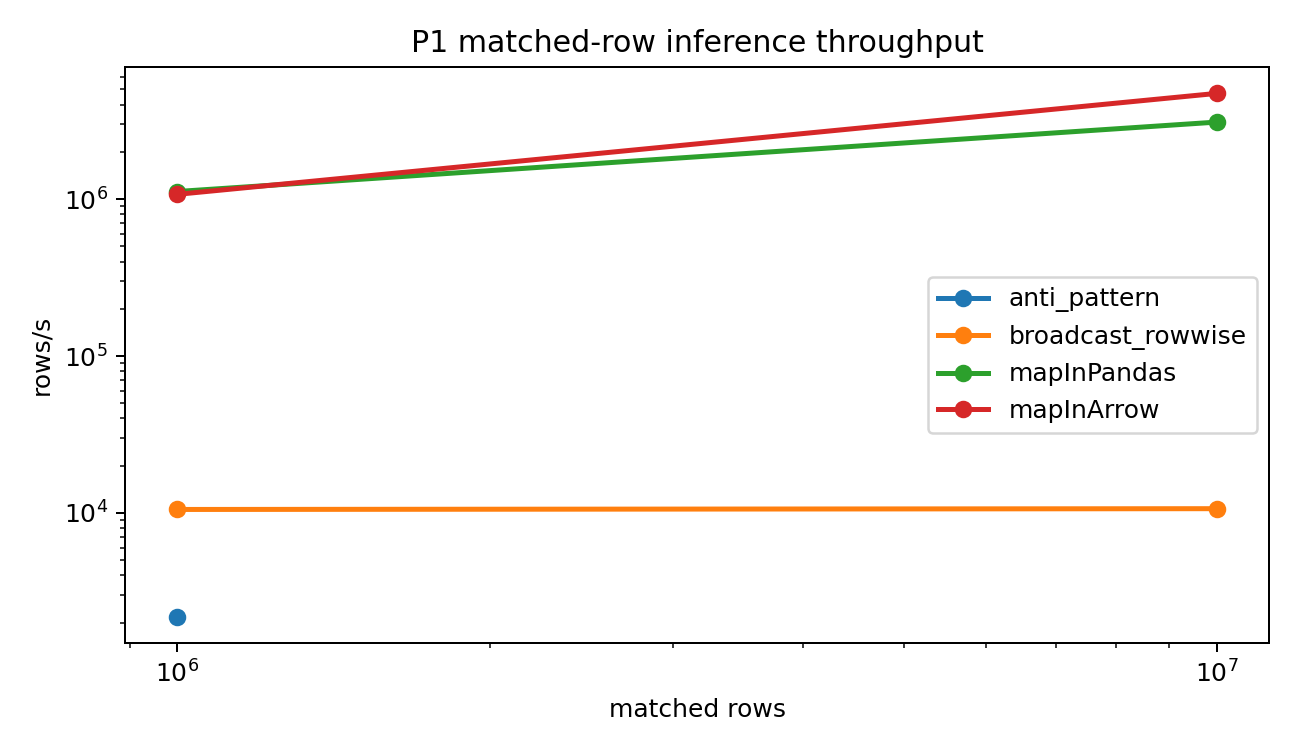}
\caption{Matched-row inference runtime and throughput for anti-pattern, broadcast-rowwise, \texttt{mapInPandas}, and \texttt{mapInArrow}.}
\label{fig:p1-matched}
\end{figure}

\begin{figure}[tbp]
\centering
\includegraphics[width=0.95\linewidth]{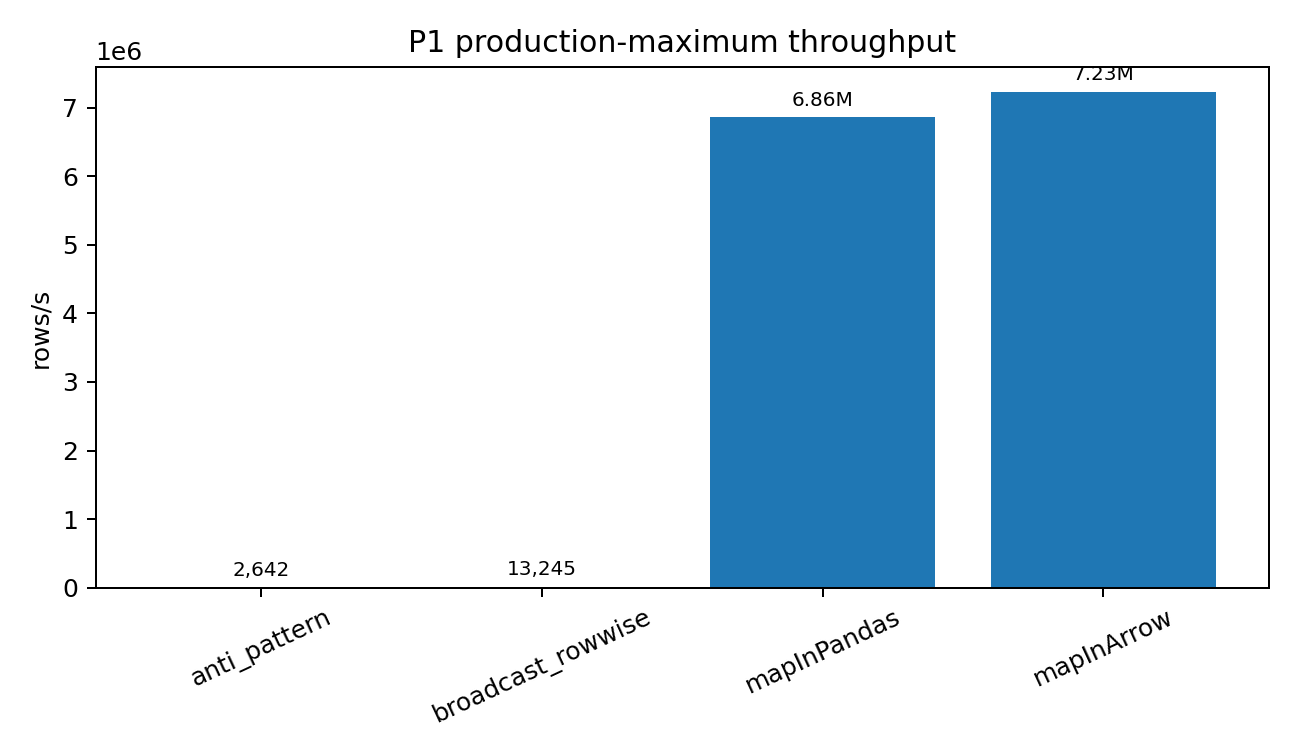}
\caption{Production-maximum inference runtime and throughput by method.}
\label{fig:p1-production}
\end{figure}

\begin{table}[tbp]
\centering
\caption{Compact P1 comparison across matched-row and production-maximum runs.}
\label{tab:p1-compact}
\small
\resizebox{\linewidth}{!}{%
\begin{tabular}{llrrrl}
\toprule
method & 1M rows/s & 10M rows/s & max rows/s & max N & role \\
\midrule
anti-pattern & 2,183 & -- & 2,643 & 250,000 & legacy worst-case \\
broadcast-rowwise & 10,522 & 10,646 & 13,245 & 1,000,000 & practical rowwise comparator \\
mapInPandas & 1.12M & 3.10M & 6.86M & 50,000,000 & vectorized pandas path \\
mapInArrow & 1.07M & 4.72M & 7.23M & 50,000,000 & vectorized Arrow path \\
\bottomrule
\end{tabular}
}
\end{table}

P2 reports measured feature-scale split-search behavior. The main paper uses no
projected points: any skipped driver baseline must record the candidate-row
count and skip reason.

The P2 result is a scale result, not a small-scale latency win. The
driver-collect reference is faster at \(\F\in\{10,50,250\}\), but the
collect-less SQL path completes for every measured point from 1240 to 124000
candidate rows with zero invalid features, while the driver reference is
skipped at \(\F=1000\) because the estimated candidate table exceeds the
100000-row safety threshold. Across those collect-less runs, the measured
driver RSS deltas stay below 1.6 MB. This is the right systems message for the
paper: driver-collect remains a reasonable small-scale reference, but the
contract-preserving collect-less path is the one that continues to operate once
candidate-table materialization stops being a sensible design.

\begin{figure}[tbp]
\centering
\includegraphics[width=0.95\linewidth]{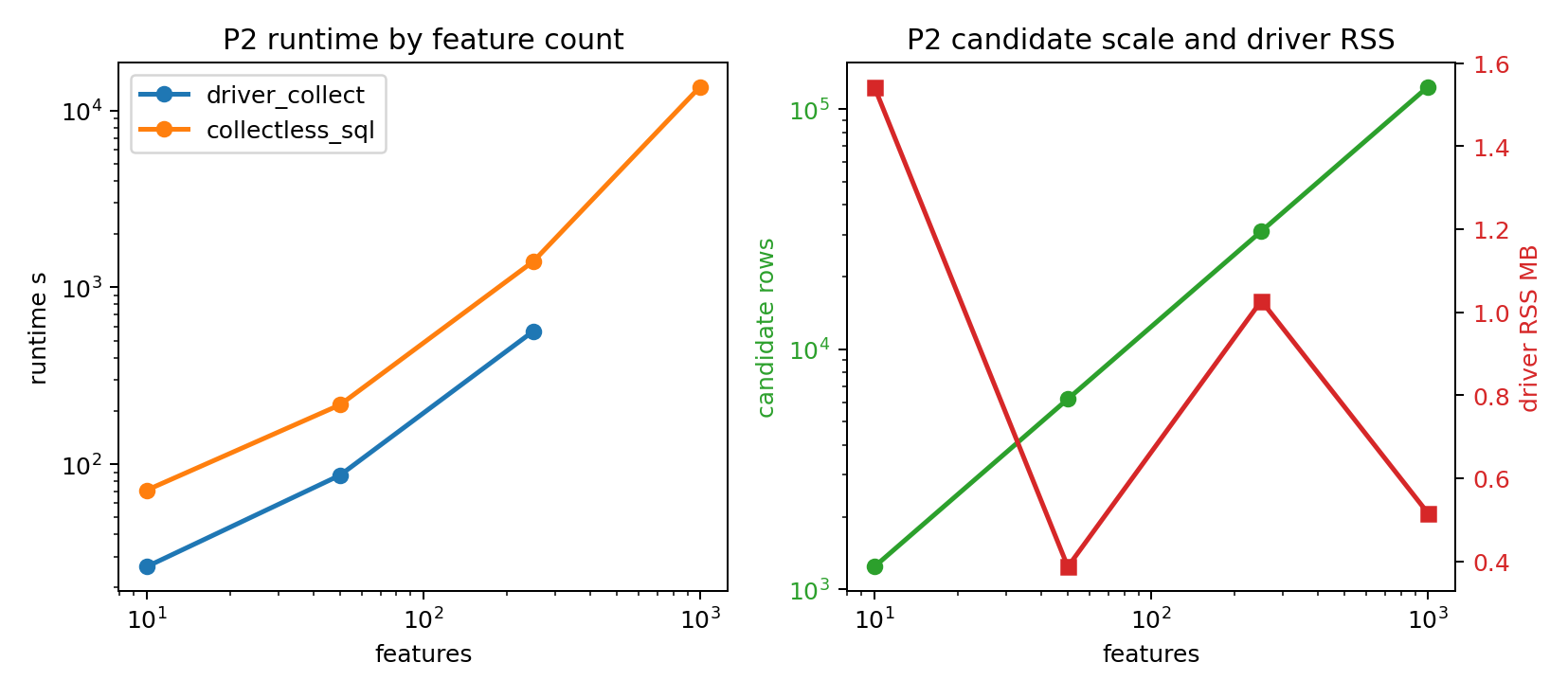}
\caption{Measured collect-less split-search scale with driver candidate rows, runtime, and driver RSS delta.}
\label{fig:p2-scale}
\end{figure}

\begin{table}[tbp]
\centering
\caption{Measured feature-scale split-search behavior.}
\label{tab:measured-split-scale}
\small
\begin{tabular}{lllllll}
\toprule
method & F & candidate rows & runtime s & driver RSS MB & scored F & status \\
\midrule
driver-collect & 10 & 1,240 & 26.24 & 0.00 & -- & ok \\
collect-less SQL & 10 & 1,240 & 71.11 & 1.54 & 10 & ok \\
driver-collect & 50 & 6,200 & 86.48 & 0.00 & -- & ok \\
collect-less SQL & 50 & 6,200 & 216.32 & 0.39 & 50 & ok \\
driver-collect & 250 & 31,000 & 565.75 & 0.00 & -- & ok \\
collect-less SQL & 250 & 31,000 & 1404.02 & 1.03 & 250 & ok \\
driver-collect & 1,000 & 124,000 & -- & -- & -- & skipped\_too\_large \\
collect-less SQL & 1,000 & 124,000 & 13647.68 & 0.52 & 1,000 & ok \\
\bottomrule
\end{tabular}
\end{table}

S2 turns the Arrow-vs-Pandas choice into a measured systems result instead of
an anecdotal observation. Across 24 workload settings varying Arrow batch size,
tree depth, number of trees, and treatment cardinality, \texttt{mapInArrow}
wins 18 settings and \texttt{mapInPandas} wins 6. Arrow dominates all eight
settings at batch size 1000 and seven of eight at 10000, but Pandas wins five
of eight at 50000, mainly on \(\T=4\) workloads and heavier model
configurations. The best observed throughputs are 8.14M rows/s for
\texttt{mapInArrow} and 7.16M rows/s for \texttt{mapInPandas}. This is the
precise claim the paper should make: Arrow is often faster, but backend choice
depends on the batch-size/model-shape regime.

\begin{figure}[tbp]
\centering
\includegraphics[width=0.95\linewidth]{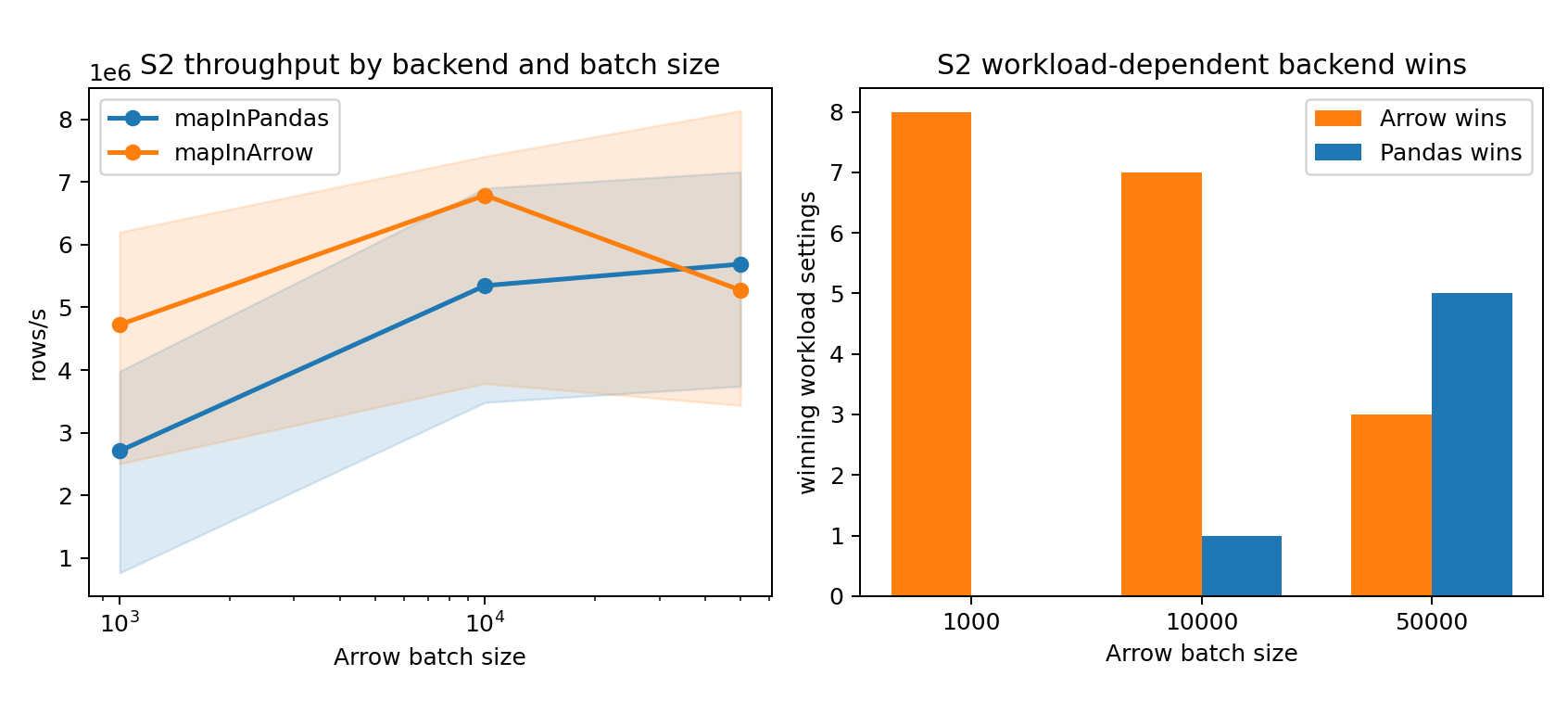}
\caption{S2 backend crossover and batch-size ablation. Left: throughput bands and medians by backend and Arrow batch size. Right: number of workload settings won by each backend at each batch size.}
\label{fig:s2-backend-crossover}
\end{figure}

\begin{table}[tbp]
\centering
\caption{S2 backend-crossover summary by Arrow batch size.}
\label{tab:s2-backend-crossover}
\small
\begin{tabular}{lllll}
\toprule
batch size & Arrow wins & Pandas wins & best Arrow rows/s & best Pandas rows/s \\
\midrule
1,000 & 8/8 & 0/8 & 6.20M & 3.98M \\
10,000 & 7/8 & 1/8 & 7.41M & 6.90M \\
50,000 & 3/8 & 5/8 & 8.14M & 7.16M \\
\bottomrule
\end{tabular}
\end{table}

\subsection{Correctness / Parity}
C1 and C2 test the primitive-level contracts. C1 compares inference score
vectors across rowwise-broadcast, \texttt{mapInPandas}, and
\texttt{mapInArrow}. C2 compares split-search outputs across three backends:
driver-collect, SQL, and \texttt{mapInPandas}. Success requires zero mismatches
above \(10^{-9}\) for inference vectors and exact best-split tuple agreement
with score equality within tolerance for split search.

All C1 backends report zero mismatches above \(10^{-9}\) and max absolute delta
0.0. All C2 backends return the same best split tuple
\texttt{("x\_boundary", 15, 0.5, "right", 0.01700393701327202)}.

\begin{table}[tbp]
\centering
\caption{Backend parity for inference and split search.}
\label{tab:backend-parity}
\small
\begin{tabular}{lllll}
\toprule
primitive & backend & mismatch rows & max delta & status \\
\midrule
inference & broadcast-rowwise & 0.0 & 0.000 & ok \\
inference & mapInPandas & 0.0 & 0.000 & ok \\
inference & mapInArrow & 0.0 & 0.000 & ok \\
split search & driver-collect & 0 & 0.000 & ok \\
split search & SQL & 0 & 0.000 & ok \\
split search & mapInPandas & 0 & 0.000 & ok \\
\bottomrule
\end{tabular}
\end{table}

\subsection{End-to-end Policy Preservation}
E1 tests a synthetic multi-depth policy-tree witness under the fixed-input
contract. E2 upgrades Hillstrom from a semantic witness to an end-to-end
policy-preservation result. Both blocks compare serialized signatures, held-out
policy vectors, treatment assignments, leaf assignments where applicable,
policy value, AUUC, and Qini.

E1 preserves the serialized signature and policy vector exactly for every
backend at depths 1 through 4, with policy agreement 1.0 throughout. E2 does
the same on Hillstrom: all backends report identical signatures, assignment
agreement 1.0, policy value 0.182109, AUUC 0.036672, Qini 0.000461, and
policy-vector max delta 0.0. E1 is a witness-style preservation result rather
than a scale result: its purpose is to show that once boundaries, feature order,
treatment order, and node expansion order are fixed, the distributed execution
paths preserve the learned tree itself. The strengthened E1 grid extends that
claim across \(\T\in\{4,8\}\) and \(\pmiss\in\{0.0,0.3\}\) at depth 2. All
12 grid rows preserve identical signatures, zero exact policy-vector
mismatches, and zero leaf mismatches across driver-collect, SQL, and
\texttt{mapInPandas}.

\begin{table}[tbp]
\centering
\caption{Synthetic end-to-end preservation witness across depth targets.}
\label{tab:e1-synthetic}
\small
\begin{tabular}{lllllll}
\toprule
depth & backends & same signature & policy agreement & policy value & AUUC & Qini \\
\midrule
1 & 6 & yes & 1.000000 & 0.200000 & -0.012500 & -0.012500 \\
2 & 6 & yes & 1.000000 & 0.200000 & -0.012500 & -0.012500 \\
3 & 6 & yes & 1.000000 & 0.200000 & -0.012500 & -0.012500 \\
4 & 6 & yes & 1.000000 & 0.200000 & -0.012500 & -0.012500 \\
\bottomrule
\end{tabular}
\end{table}

\begin{table}[tbp]
\centering
\caption{Extended E1 preservation grid across treatment cardinality and missingness.}
\label{tab:e1-extended}
\small
\resizebox{\linewidth}{!}{%
\begin{tabular}{lllllllll}
\toprule
T & pmiss & backends & same signature & max policy mismatches & max leaf mismatches & policy value & AUUC & Qini \\
\midrule
4 & 0.0 & 3 & yes & 0 & 0 & 0.200000 & 0.099375 & -0.000625 \\
4 & 0.3 & 3 & yes & 0 & 0 & 0.200000 & 0.099375 & -0.000625 \\
8 & 0.0 & 3 & yes & 0 & 0 & 0.200000 & 0.104375 & 0.004375 \\
8 & 0.3 & 3 & yes & 0 & 0 & 0.200000 & 0.104375 & 0.004375 \\
\bottomrule
\end{tabular}
}
\end{table}

\begin{figure}[tbp]
\centering
\includegraphics[width=0.95\linewidth]{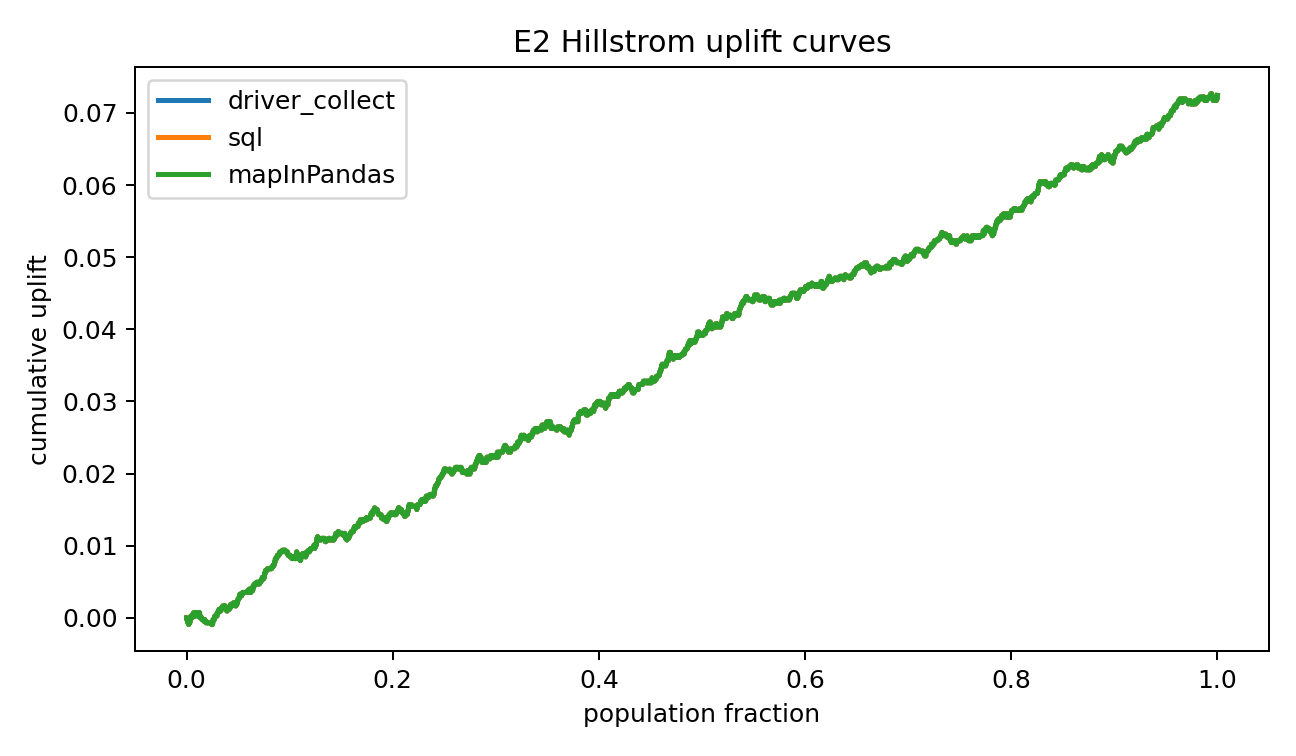}
\caption{Hillstrom AUUC and Qini curves for the single-node reference and distributed toolkit paths.}
\label{fig:e2-hillstrom-curves}
\end{figure}

\begin{table}[tbp]
\centering
\caption{Hillstrom end-to-end policy preservation and utility metrics.}
\label{tab:hillstrom-utility}
\small
\begin{tabular}{lllllll}
\toprule
backend & same signature & assignment agreement & policy value & AUUC & Qini & max delta \\
\midrule
driver-collect & yes & 1.000000 & 0.182109 & 0.036672 & 0.000461 & 0.000 \\
SQL & yes & 1.000000 & 0.182109 & 0.036672 & 0.000461 & 0.000 \\
mapInPandas & yes & 1.000000 & 0.182109 & 0.036672 & 0.000461 & 0.000 \\
\bottomrule
\end{tabular}
\end{table}

\subsection{Robustness Under Perturbation}
S1 asks whether the contract stays exact under realistic Spark perturbations.
It varies repartition counts, shuffles rows before lock, uses
\texttt{coalesce}, and toggles intra-partition sorting. F3 asks the same
question for missingness semantics. It sweeps missingness rates 0.0 through
0.5, both NULL and NaN encodings, and missingness concentrated in the control
arm, treated arms, or the positive-outcome subgroup.

S1 gives the right semantics story. Once the fixed-input lock is enforced, all
six perturbation variants preserve the serialized signature exactly, with zero
exact policy-vector mismatches and zero leaf mismatches. Before lock, all six
variants drift, with signature mismatches in every case, exact policy-vector
drift in every case, and leaf drift in three of six variants. That is strong
evidence that the contract is doing real work rather than merely documenting an
already-stable pipeline.

F3 is equally clean. Across 72 D1 parity checks, 72 D2 parity checks, and 54
representative end-to-end witness checks, every contract-preserving row is
exact: no inference mismatches, no split-tuple mismatches, no signature drift,
no exact policy-vector drift, and no leaf drift. The paper can therefore say
that the toolkit is robust to this missingness stress grid under the fixed-input
contract, not merely on one nominal data distribution.

\begin{table}[tbp]
\centering
\caption{S1 partition/order perturbation robustness before and after manifest lock.}
\label{tab:s1-robustness}
\small
\begin{tabular}{llllll}
\toprule
state & variants & same signature & exact policy drift variants & leaf drift variants & max vector delta \\
\midrule
after lock & 6 & 6/6 & 0/6 & 0/6 & 0.000000 \\
before lock & 6 & 0/6 & 6/6 & 3/6 & 0.314286 \\
\bottomrule
\end{tabular}
\end{table}

\begin{table}[tbp]
\centering
\caption{F3 missingness-stress summary across primitive and end-to-end exactness checks.}
\label{tab:f3-missingness}
\small
\resizebox{\linewidth}{!}{%
\begin{tabular}{p{0.18\linewidth}r r p{0.46\linewidth} p{0.14\linewidth}}
\toprule
component & cases & exact cases & grid & worst observed delta \\
\midrule
D1 parity & 72 & 72 & pmiss in \{0, 0.1, 0.3, 0.5\}; NULL and NaN; control/treatment/positive focus & 0.000 \\
D2 parity & 72 & 72 & pmiss in \{0, 0.1, 0.3, 0.5\}; NULL and NaN; control/treatment/positive focus & exact tuple match \\
end-to-end witness & 54 & 54 & pmiss in \{0, 0.1, 0.3, 0.5\}; NULL and NaN; control/treatment/positive focus & 0.000 \\
\bottomrule
\end{tabular}
}
\end{table}

\subsection{Failure / Stress Results}
F1 records which intentionally naive variants drift and which contract rule
fixes the drift. F2 isolates boundary sensitivity by comparing frozen
boundaries with independently recomputed approximate boundaries on the same
data and seed. The expected outcome is equality under frozen boundaries and
visible drift under recomputation; if the frozen-boundary path diverges, the
semantics must be fixed before polishing the paper.

All four naive F1 variants fail at least once. \texttt{naive\_first\_seen\_control}
drifts in 34 cases, \texttt{naive\_no\_total\_order} in 29,
\texttt{naive\_sparse\_omit} in 30, and \texttt{naive\_implicit\_missing} in
3. The strongest evidence comes from \texttt{x\_tie}, \texttt{x\_boundary},
\texttt{x\_miss}, and \texttt{x\_sparse}; \texttt{x\_control} mainly acts as a
scope-boundary witness where the contract rejects the candidate set outright.
For \texttt{unstable argmax}, the drift rows are equal-score ties: the winning
candidate identity changes even though the score itself does not, so the largest
score delta remains 0.0 while the selected split still drifts.
The most operationally dangerous naive variant is \texttt{naive\_sparse\_omit},
because 30 of its failures come from accepting a split after the contract path
has already rejected the candidate set as invalid. That behavior is exactly why
candidate validity belongs in the semantic contract rather than in ad hoc
backend-specific logic.
In F2, the fixed-boundary path returns a valid split, while the independently
recomputed approximate-quantile path produces no valid D2.2 candidate on the
same data and seed. The stronger witness result shows why that matters:
changing the boundary from \texttt{0.5} to \texttt{0.4999999999} changes the
serialized tree signature and 99996 held-out policy vectors and leaf
assignments, even though the top-treatment assignment agreement remains 1.0.
The AUUC and Qini values also shift by about 0.0078. So the paper should not
frame boundary recomputation as a harmless numerical detail. It changes the
learned policy witness materially, even when the top treatment label stays the
same on most rows.

\begin{figure}[tbp]
\centering
\includegraphics[width=0.95\linewidth]{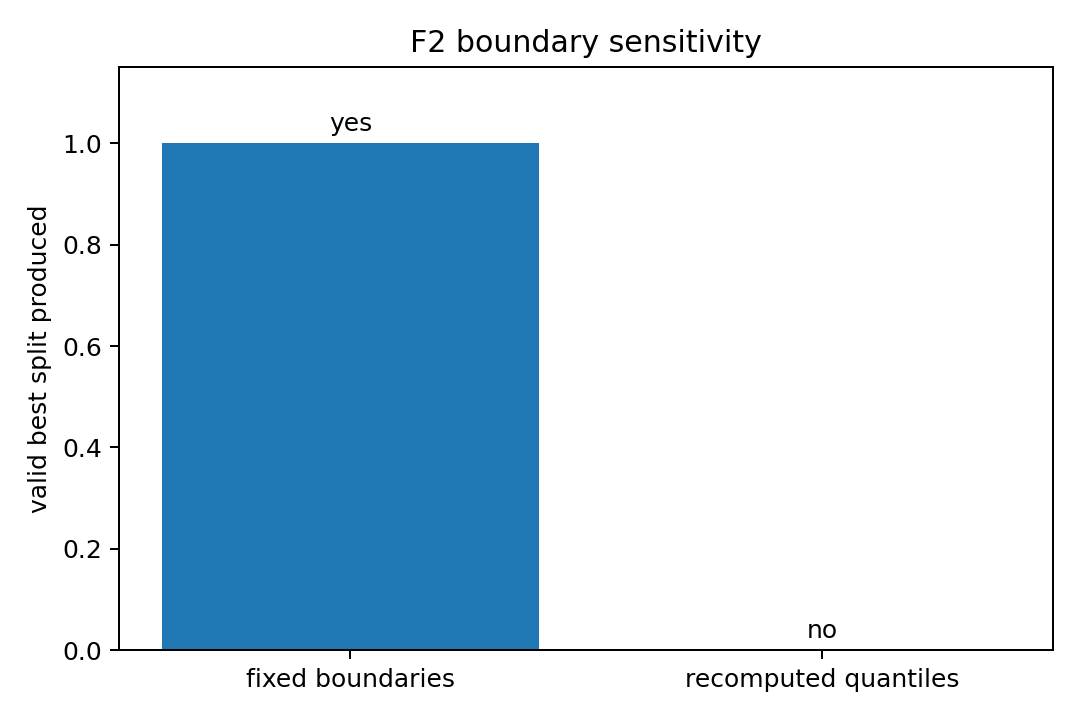}
\caption{Fixed-boundary versus independently recomputed-boundary valid-split availability on the same data and seed.}
\label{fig:f2-boundary}
\end{figure}

\begin{table}[tbp]
\centering
\caption{F2 boundary-witness drift under frozen versus independently recomputed boundaries.}
\label{tab:f2-boundary-witness}
\small
\resizebox{\linewidth}{!}{%
\begin{tabular}{l l l l r r r r r}
\toprule
fixed & recomputed & signature drift & assignment agreement & policy mismatches & leaf mismatches & max delta & AUUC abs delta & Qini abs delta \\
\midrule
b=[0.5] & b=[0.4999999999] & yes & 1.000000 & 99,996 & 99,996 & 0.052229 & 0.007815 & 0.007815 \\
\bottomrule
\end{tabular}
}
\end{table}

\subsection{Cross-Experiment Summary}
Table~\ref{tab:key-results} collects the results that anchor the paper's claims.
The speed evidence comes from P1 and S2, the scale evidence from P2, the
primitive contracts from C1/C2 and F3, the end-to-end preservation evidence
from E1/E2 and S1, and the failure semantics from F1/F2. Grouping the evidence
this way makes the paper's identity explicit: one toolkit, two primitives, one
semantic contract, and one coherent validation program.

\begin{table}[tbp]
\centering
\caption{Cross-experiment summary of the validated semantic-contract results.}
\label{tab:key-results}
\small
\begin{tabular}{p{0.25\linewidth}p{0.65\linewidth}}
\toprule
result & validated outcome \\
\midrule
P1 practical comparator at 10M & mapInPandas 290.8x and mapInArrow 443.1x over broadcast\_rowwise \\
P1 production maximum & mapInArrow 7.23M rows/s; mapInPandas 6.86M rows/s \\
P2 largest measured feature count & collectless\_sql ok at F=1,000 (124,000 candidates); driver\_collect skipped as too large \\
C1 and C2 primitive parity & zero inference mismatches above 1e-9 and exact best-split tuple agreement \\
E1 synthetic witness & depths 1 through 4 preserve identical signatures and policy vectors across all backends \\
E1 extended grid & T=\{4,8\} and pmiss=\{0.0,0.3\} preserve zero policy and leaf mismatches \\
E2 Hillstrom preservation & assignment agreement 1.000000, policy value 0.182109 \\
S1 partition/order robustness & after lock 6/6 same signature; before lock 0/6 \\
F1 observed naive drifts & first-seen control 34, unstable argmax 29, sparse omission 30, implicit missing 3 \\
F2 boundary sensitivity & fixed boundaries yield a valid split; recomputed approximate quantiles do not \\
F3 missingness stress & D1 72/72 exact, D2 72/72 exact, end-to-end 54/54 exact \\
S2 backend crossover & mapInArrow wins 18/24 workload settings; mapInPandas wins 6/24 \\
\bottomrule
\end{tabular}
\end{table}

\section{Discussion and Limitations}
Spark Policy Toolkit is not a replacement for optimized supervised tree
systems, and it does not claim arbitrary determinism for arbitrary Spark jobs.
Its role is narrower and more operational: preserve the meaning of a custom
policy-learning pipeline while moving the expensive parts of inference and
split search into Spark-native execution. In that sense it belongs closer to
the correctness and semantic-faithfulness lineage represented by
\texttt{mlf-core} and AERIFY than to estimator-centric uplift packages or
distributed causal estimators \cite{heumos2023mlfcore,zulkifli2025aerify}.

The fixed-input contract is also a limitation. The toolkit assumes the
preprocessing manifest, treatment vocabulary, feature order, and split
boundaries are locked before comparing backends. Approximate quantile
boundaries may be useful for exploration, but independently recomputing them in
different paths is outside the semantic guarantee.

The split-search scale result should also be read precisely. P2 does not claim
that collect-less SQL is the fastest path at small feature counts; it shows
that the contract-preserving path remains operational when driver-side
candidate-table materialization stops being a reasonable execution strategy.

Selected legacy measurements remain useful as appendix-only systems context.
The older E4 split-search sweeps show a clean bin-count-dependent crossover
between SQL and \texttt{mapInPandas}, and the refreshed 8-worker versus
40-worker E1/E2 reruns show the expected throughput/runtime penalty at smaller
cluster size. Those measurements strengthen interpretation of backend choice
and cluster sensitivity, but they are intentionally kept out of the core claim
path because they were not run as part of the frozen semantic-contract bundle
used for P1/P2/C1/C2/E1/E2/F1/F2.

The older Spark UI measurements are also useful for interpretation. Across the
refreshed 40-worker and 8-worker inference-cost runs, GC ratios remain low
(below 1.21\%; see Appendix~A), while executor CPU time grows sharply for the
rowwise paths. That pattern supports the paper's central reading of the
serialization tax: the bottleneck is dominated by Python-side and executor CPU
work, not by a GC-driven pathology.

The current evidence is also intentionally scoped. The paper now shows exact
fixed-input preservation on the synthetic witness and on Hillstrom, exactness
under a missingness stress grid, and stability under six realistic
repartition/coalesce/shuffle perturbations once the manifest lock is enforced.
It therefore includes one public real-data end-to-end preservation case, not a
broad multi-dataset real-data survey. It still does not claim arbitrary
determinism outside that lock, arbitrary stability under independently
recomputed boundaries, or broad public-dataset coverage. Extending the same
end-to-end witness discipline to additional public datasets remains the most
obvious next breadth step.

\section{Conclusion}
Spark Policy Toolkit is a semantics-governed execution layer for policy
learning in Spark. Vectorized inference and collect-less split search improve
scalability, but their purpose is larger than raw speed: they preserve
inference outputs, split decisions, and learned policy behavior under an
explicit fixed-input contract. The validated run supports two clean claims:
vectorized inference delivers the large runtime gains, and collect-less split
search carries the semantic contract to larger candidate tables without
driver-side materialization. The strengthened final-round evidence adds three
important refinements: backend choice is workload-dependent, missingness stress
does not break the contract-preserving paths, and realistic Spark perturbations
remain exact once the fixed-input lock is enforced. The paper's core message is
semantics and scalability together.

\appendix

\section{Legacy Supplemental Systems Context}
This appendix reuses a narrow subset of older measurements that answer systems
questions not directly covered by the frozen semantic-contract bundle. These
artifacts are supplemental only: they do not support the core semantic
preservation claims in the main text.

\subsection{Legacy Split-Search Backend Crossover}
The older E4 sweep remains useful because it isolates a backend-choice question
that the current semantic bundle does not measure directly. On both the
40-worker and 8-worker clusters, \texttt{mapInPandas} is slightly faster at 16
bins, the two backends are close around 32 bins, and SQL is faster by 64 and
128 bins. That is consistent with the main paper's broader message that
backend choice is workload-dependent rather than universal.

\begin{figure}[tbp]
\centering
\includegraphics[width=0.95\linewidth]{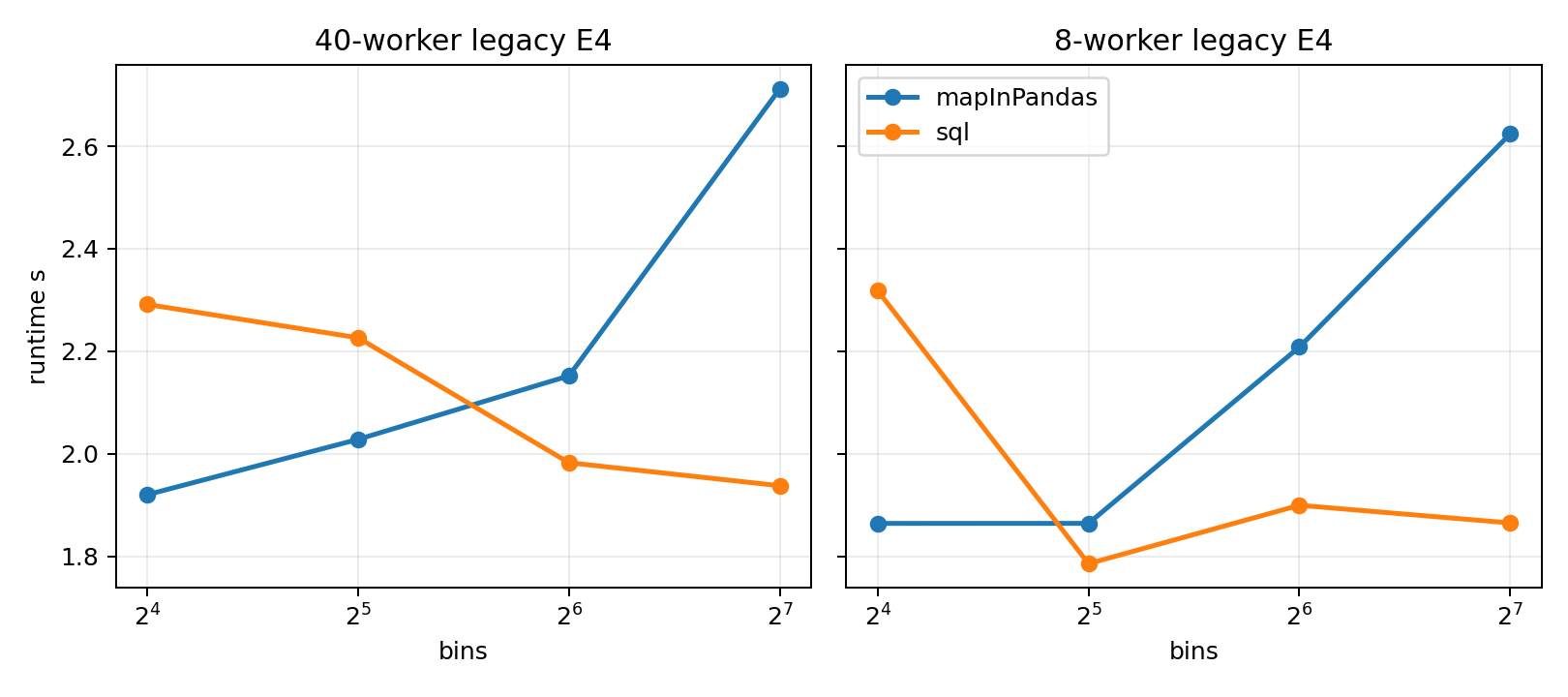}
\caption{Appendix-only legacy E4 backend crossover from the older paper notebook family. The panels use the 40-worker and 8-worker legacy runs and show a bin-count-dependent crossover between SQL and \texttt{mapInPandas}.}
\label{fig:appendix-legacy-e4}
\end{figure}

\subsection{Cluster-Size Sensitivity and CPU/GC Context}
The refreshed legacy E1/E2 reruns provide clean 40-worker versus 8-worker
context for inference behavior. They are not part of the core semantic bundle,
but they are useful for interpreting how the same inference workload scales
with cluster size. The same reruns also include Spark UI CPU and GC summaries,
which help explain the rowwise bottleneck without turning manual profiling into
a main-text claim.

\begin{table}[tbp]
\centering
\caption{Refreshed legacy E1/E2 appendix-only cluster-size context from the 40-worker and 8-worker reruns.}
\label{tab:appendix-cluster-sensitivity}
\small
\begin{tabular}{p{0.30\linewidth}p{0.17\linewidth}rrrr}
\toprule
workload & method & 40-worker & 8-worker & 8/40 ratio \\
\midrule
E1 10M inference & mapInPandas & 3.80M & 2.47M & 0.65 \\
E1 10M inference & mapInArrow & 3.10M & 2.22M & 0.72 \\
E2 50M runtime & mapInArrow & 7.10 & 23.37 & 3.29 \\
E2 1M runtime & broadcast\_rowwise & 75.71 & 241.90 & 3.20 \\
\bottomrule
\end{tabular}
\end{table}

\begin{table}[tbp]
\centering
\caption{Legacy Spark UI CPU and GC measurements for the refreshed inference-cost runs. These measurements are appendix-only interpretive support, not core semantic evidence.}
\label{tab:appendix-cpu-gc}
\small
\begin{tabular}{l l r r r}
\toprule
cluster & method & CPU s & GC s & GC ratio \\
\midrule
40-worker & mapInArrow & 962 & 3.424 & 0.356\% \\
40-worker & broadcast\_rowwise & 11478 & 0.304 & 0.003\% \\
40-worker & anti\_pattern & 12672 & 6.328 & 0.050\% \\
8-worker & mapInArrow & 1930 & 23.305 & 1.208\% \\
8-worker & broadcast\_rowwise & 19914 & 0.208 & 0.001\% \\
8-worker & anti\_pattern & 22164 & 94.952 & 0.428\% \\
\bottomrule
\end{tabular}
\end{table}

\bibliographystyle{plainnat}
\bibliography{bib/references}

\end{document}